# TWO YEARS OF DETECTING DM OBJECTS - THE SOLAR SYSTEM MEMBERS

E. M. DROBYSHEVSKI, M. E. DROBYSHEVSKI, T. Yu. IZMODENOVA, and D. S. TELNOV

*Ioffe Physico-Technical Institute, Russian Academy of Sciences, 194021 St.Petersburg, Russia*
*(E-mail: emdrob@pop.ioffe.rssi ru)*

With a probability >99% there are grounds to believe that our works on detection of the Dark Electric Matter Objects (daemons), which were launched in 1996, are crowned with success. The daemons are the relic elementary Planckian black holes ($m \approx 30$ μg) carrying stable electric charge of $Z = 10e$. During the last two years, the detector made of two horizontal ZnS(Ag) screens of 1 m$^2$ area has been recording the correlated time-shifted scintillations corresponding to flux $f_\oplus \sim 10^{-5}$ m$^{-2}$s$^{-1}$ of extraordinary penetrating nuclear-active particles which moved both down and upward with a velocity of only ~5-30 km/s. The flux experiences seasonal variations with maxima supposedly corresponding to the Earth transition through shadow and anti-shadow created by the Sun in its motion relative the Galaxy disk daemon population. An accumulation of negative daemons, which stimulate the proton decay in ~1 μs, inside the Earth and the Sun is capable of explaining a lot of previously non-understandable facts.

1 INTRODUCTION. GENERAL IDEOLOGY AND HISTORY OF OUR SEARCH

We started from a working hypothesis that the DM of the Galactic disk consists of electrically charged Planckian elementary black holes ($m = 3 \cdot 10^{-5}$ g, $r_g = 2 \cdot 10^{-33}$ cm). The charge of these DArk Electric Matter Objects, daemons, may constitute $Z = 10e$ (Markov, 1965).
   Having a large charge, they are slowed down fairly efficiently in their passage through the Sun, to become captured into strongly elongated orbits with perihelia inside it. Because of the resistance offered by the solar matter, these orbits contract to move into the Sun, with daemons building up in the latter. In 4.5 Gyr, their number could reach ~$2.4 \cdot 10^{30}$ (Drobyshevski, 1996). If a daemon captured in such a heliocentric orbit passes through the Earth's sphere of influence, it will be deflected, and its perihelion will move out of the Sun, to produce a fairly stable population in strongly elongated, Earth-crossing heliocentric orbits (SEECHOs). Optimistic estimates yield for its concentration a value ~$10^3$–$10^5$ larger than that of daemons in the Galactic disk (Drobyshevski, 1997). Perturbations by the Earth transfer part of the daemons from here to near-Earth, almost circular heliocentric orbits (NEACHOs), whence they can fall on the Earth with a velocity ~11−15 km/s. At a velocity close to 11.2 km/s, the slowing down rendered by the Earth's material may turn out sufficient to transfer a daemon from a NEACHO to a geocentric Earth-surface-crossing orbit (GESCO). GESCOs contract, and daemons go into the Earth. Thus, we see that the Solar system should harbor more than one population of daemons moving with velocities $V < 50$ km/s relative to the Earth. Therefore, in contrast to all other scientific teams that made a search for objects of the Galactic halo DM with $V \sim$ 200−300 km/s their goal, we decided to look for objects, more specifically, daemons, with $V < 50$ km/s.
   The large charge and the large mass make daemons nuclear-active particles. The possibility of using negative daemons for catalysis of light nuclei (up to O and F) appears obvious (Drobyshevski, 1997). This is why we made an attempt to use this property for their detection. One might expect that when a daemon passed through Li or Be with $V \sim 50$ km/s, up to $10^2$–$10^3$ erg/cm would be released along its trajectory. Expansion of heated material would generate a sound wave, whose detection should permit one to determine the daemon velocity and trajectory. However, experiments undertaken with Li (end of 1996 to beginning of 1998) revealed a strong damping of ultrasound of 20–30 MHz frequency along a path of only ~1 cm. An analysis of the possibility of using Be showed the corresponding frequency to be 500–600 MHz. Its detection would require employing a fairly sophisticated technique. Therefore, in March, 1998, we terminated the experiments on acoustic detection of daemons.

As a daemon enters or exits Be, part of the $^{18}$O nuclei forming in the catalytic fusion of $^9$Be nuclei should escape outward. If the surfaces of Be plates are coated with a scintillator, e.g., ZnS(Ag), the shift of the signals initiated in them would indicate the passage of a daemon. However, an exposure of Be plates 0.12 m$^2$ in area carried out in June, 1999, for 300 h did not yield any sensible result, whence we concluded that the flux of the SEECHO daemons $f_\oplus < 3 \cdot 10^{-4}$ m$^{-2}$s$^{-1}$ (Drobyshevski, 2000b). Extending the exposure to 500 h did not change anything.

The experiments with Be, which in our case contained up to 0.1 at.% impurities with $Z_n > 10$, suggested a possible role of "poisoning" of the catalyst as a result of the daemon capturing heavy nuclei. As follows from an analysis of Solar energetics (which was not quite correct, as we understand it now), a daemon should free itself of a heavy nucleus in a time $t_{ex} \sim 10^{-7}$–$10^{-6}$ s because of the disintegration of the daemon-containing protons (Drobyshevski, 2000a,b).

This led us to a radically new ideology of their detection, which is based on the assumed fast daemon-stimulated proton decay.

The system constructed in October–November, 1999, consisted of four moduli. Each module had two parallel, transparent 4-mm thick polystyrene screens coated on the underside with ZnS(Ag) and arranged at the center of a cubic tinned-sheet case 51 cm on a side. The case was covered on the front with black paper (for more details, see Drobyshevski, 2002b). The screens were mutually light-isolated and disposed at a distance of 7 cm from one another. Each screen was viewed by a separate PM tube, whose output signals were fed to a double-trace oscilloscope. It was assumed that on passing through the first screen and on capturing a Zn or S nucleus in the scintillator on the way, the daemon, in crossing the 7-cm gap, while releasing the nucleus because of the proton decay, would excite a scintillation, which would trigger the oscilloscope. On entering the second ZnS(Ag) layer, it would capture another nucleus and, on initiating its disintegration, would excite a scintillation to be detected by the second PM tube connected to the second trace of the oscilloscope. The magnitude of the signal shift $\Delta t$ would then provide a judgement of the daemon velocity $V$.

Initially, we attempted to detect the SEECHO population, for which purpose the system was continuously oriented at the desired point on the celestial sphere (its coordinates for different dates were kindly calculated by Yu.D.Medvedev). Round-the-clock exposure run during December 1999–January 2000 did not, however, produce any definite result.

In February, 2000, the system was oriented horizontally and was switched on only in day-time. An analysis of the distribution $N(\Delta t)$ of the upper and bottom PM tube signals in the relative time shift $\Delta t$ exhibited a certain indication of statistically significant deviations from the constant background originating from natural radioactivity and intrinsic PM noise. After this, the system in its horizontal arrangement was switched on February 24, 2000, to round-the-clock operation. By now, this operation has yielded the discovery of negative daemons and of their slow populations in the Solar system, and revealed the main modes of daemon interaction with matter.

2 KEY STAGES IN THE EXPERIMENT

*2.1. The first statistically significant results*

The March, 2000, experiment revealed within the interval $-100 < \Delta t < +100$ μs an $N(\Delta t)$ distribution which deviated from $N$ = const, by the $c^2$ criterion, with a probability of 99%. This distribution, for a total number of 417 events, exhibited a distinct enough maximum in the interval $+20 < \Delta t < 40$ μs. It exceeded the average level of 41.7 events per bin by 2.7$s$ [where $s = (41.7)^{1/2}$]. Its presence aroused some optimism, albeit not without a trace of perplexity, because the velocity derived from the base length of 7 cm constituted only 3−4 km/s.

Because of the extreme potential significance of this result, we carried out a large number of check experiments to reveal possible artefacts or faults in the operation of our simple equipment. None were revealed. Simultaneously the detector itself was modified. For instance, experiments with tinned iron sheets placed between the scintillators were performed (Drobyshevski, 2000c). Although this arrangement produced some effect, we did not succeed in drawing any definite conclusions therefrom. It should be stressed that we have not succeeded in finding any relation between the events responsible for the maximum in the interval $20 < \Delta t < 40$ μs and cosmic rays. To enhance the detector sensitivity, the thickness of the ZnS(Ag) layer was increased from $h \approx 3.5$ mg/cm$^2$ to 5−6 mg/cm$^2$ (which, as we understand it now, should not be done: see below). The PM tube output circuit was simplified by replacing the inductance in the anode load with a 30-kΩ resistance. The software intended for triggering the oscilloscopes after each event and for signal processing was improved. The irreproducibility of some experiments, which were carried out one after another in seemingly the same conditions (although slight changes were continuously introduced), was deeply annoying.

*2.2. Revealing the important role of Heavy Particle Scintillations ( HPSs)*

The experiments revealed scintillations of two types. Belonging to the first of them are long events, with a smooth maximum, 2−2.5 μs after the beginning. This shape is characteristic of the scintillations produced by α-particles (Heavy Particle Scintillations, HPSs). The short scintillations have a maximum forming immediately after the trigger (<0.5 μs). Their decay is determined only by the parameters of the PM tube−oscilloscope circuit. This shape is typically observed in scintillations excited by electrons, as well as in those occurring without any time shift ($\Delta t = 0$) on both traces, and sometimes in several modules simultaneously. The zero-shift signals are caused by muons and electron avalanches of cosmic rays, whose primary particles, judging from the frequency of their occurrence, have an energy of $10^{11}$−$10^{13}$ eV. These events were used by us to determine the relative sensitivity of the channels but were disregarded in drawing $N(\Delta t)$ histograms. The pure PM noise signals have the same shape. Therefore we call them Noise-Like Scintillations (NLSs).

An important observation was made in October, 2000. If one takes into account only the HPS events from the top PM tube, the significance of deviation of $N(\Delta t)$ from a constant background increased, despite the general decrease in the number of points in the March experiment to 231, to 1 - $a$ = 99.7%, and the maximum at $20 < \Delta t < 40$ μs was found to exceed the average level of 23.1 event/bin by 3.7$s$ [where $s = (23.1)^{1/2}$]. Therefore we included in our analysis subsequently only the events triggered by an HPS in the upper scintillator.

*2.3. Using the HPS shape for detection of the NEACHO, GESCO, and, possibly, SEECHO populations*

In November, 2001, we became confident that the system operated without malfunctions, and that the irreproducibility of the monthly averages was caused apparently by seasonal variations of the daemon flux mainly. It was decided to conduct the experiment on the round-the-clock basis, without changing the system parameters. These observations (April−June, 2001) did indeed reveal a strong variability of $N(\Delta t)$ from one month to another.

By this time, we not only understood the reason for the significance of the HPSs in the top scintillator but realized also the possibility of using the specific features in the HPS shape, which depend on the actual direction of daemon propagation, for diagnostics. The point is that the daemon binding energy to a nucleus, $W = 1.8ZZ_n A_n^{-1/3}$ MeV, is 130 MeV for Zn. Therefore, the first to be emitted in the capture of a nucleus by the daemon are the atomic Auger electrons with an energy of up to ~1 MeV. They excite NLSs. The latter, however, become immediately superposed upon by the HPSs, which are created by the nucleons and their clusters ejected by the excited nucleus. As a result, the oscilloscopic traces of scintillations produced by fast enough daemons propagating downward [first through the polystyrene, and after that, through the ZnS(Ag) layer], should have a large area $S$ (normalized to the amplitude), i.e., they should be wider than those generated by the upward flying daemons, where part of the nucleons are ejected already in polystyrene and, hence, do not reach the ZnS(Ag) layer. This effect is indeed observed, if one separates the "wide" from "narrow" scintillations and constructs for them separate distributions, $N_w(\Delta t)$ and $N_n(\Delta t)$. The number of events per $0 < \Delta t < 20$ μs bin for $N_w(\Delta t)$ invariably exceeds that for $-20 < \Delta t < 0$ μs, and vice versa.

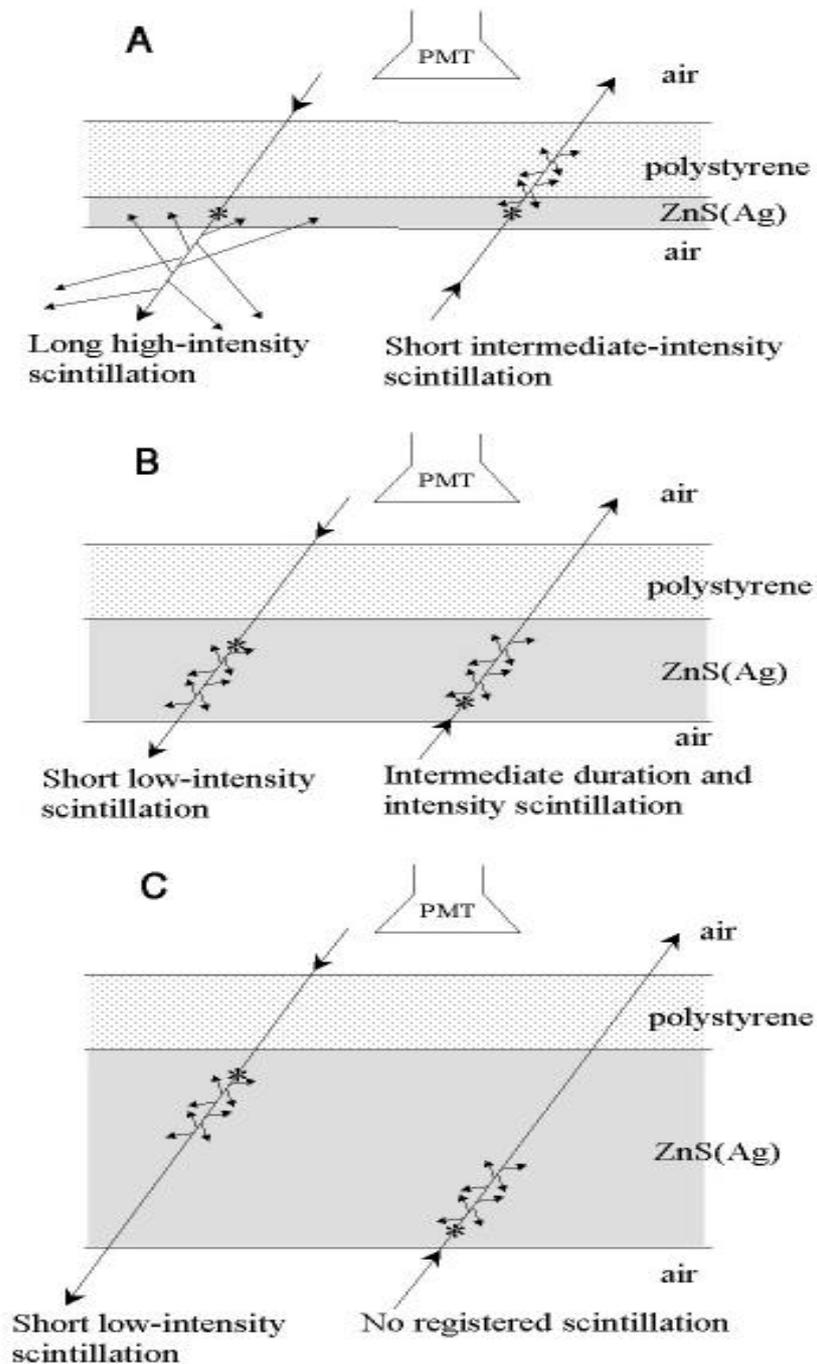

Figure 1. Parameters of the scintillations produced by slow ($V \approx 10$ km/s) daemons in ZnS(Ag), an intermediate atomic-weight, low-transparent scintillator (* - point of a nucleus capture and emission of Auger electrons; after this point, the excited captured nucleus emits about ten nucleons).
A - thin-layer scintillator ($h \leq 3.5$ mg/cm$^2$); B - thick-layer scintillator ($h \approx 6\text{-}8$ mg/cm$^2$); C – very thick-layer scintillator ($h \geq 10$ mg/cm$^2$).

While the situation becomes more complicated for lower velocities and larger ZnS(Ag) thicknesses $h$, nevertheless, it allows interpretation along the same lines (see Fig. 1). We note also that because our ZnS(Ag) powder consists of light-scattering grains differing in size (with an average size ≈12 μm), the light transmission is not related to $h$ through a simple exponential (see also Birks, 1964). From our calibration, it can be accepted as a rough estimate that for $h = 2.5$ mg/cm$^2$ the layer transmits 30% of the incident light, and for 5 mg/cm$^2$, 20%. One should bear in mind that for $h < 2$ mg/cm$^2$, ZnS(Ag) grains do not cover completely the screen surface, whereas a substantial increase of $h$ increases the number of background HPSs due to the impurities present in ZnS(Ag).

Realization of the role of the HPSs and of the dependence of their shape on the magnitude and direction of velocity and on the ZnS(Ag) thickness permitted us to construct a noncontradictory scenario of the sequence of the events involved.

On entering the detector on its way down, the daemon captures a Zn (or S) nucleus with emission of Auger electrons and nucleons (and of their clusters) from the excited nucleus. A "wide" HPS is produced. After this, a successive decay of daemon-containing protons starts in the remainder of the nucleus. As long as the daemon/remainder complex is positively charged, (re)capture of another nucleus is not likely. Therefore, the daemon crosses the bottom ZnS(Ag) layer without exciting a scintillation. However, when approaching the bottom lid of the case with the velocity $V \sim 10$ km/s, the daemon is already capable of capturing here a Sn (or Fe) nucleus with the emission of electrons at first via the Auger process and then by means of internal conversion of the excited nucleus energy to the refilling (i.e. anew captured by the nucleus from the ambient metal) electrons. Part of them, on traversing the distance of ~22 cm to the bottom scintillator, excite in it an NLS shifted by $\Delta t$ (22 cm of air are impenetrable for the nucleons evaporating from the nucleus), and this event is detected.

When moving upward with a velocity $V \sim 10$ km/s, the daemon is poisoned by a Sn nucleus and has not time enough to "digest" it before entering the top ZnS(Ag) layer. This accounts for the absence of a maximum in the interval $-40 < \Delta t < -20$ μs in the March, 2000, distribution $N(\Delta t)$. For $V > 10$ km/s, it crosses the thin (2 μm) Sn layer without capturing Sn, but captures Fe. For $10 < V < 15–20$ km/s, the Fe nucleus can be digested by the daemon before it strikes the top ZnS(Ag) layer, and then the upward crossing of the detector by the daemon will be registered.

Thus, the distance of 29 cm between the top ZnS(Ag) layer and the bottom lid of the case is the base length in our experiment. For $V > 40$ km/s, a detector of our dimensions becomes "transparent" for daemons. For $V < 3–5$ km/s, the base length would be the separation between the top and bottom scintillators (as this was supposed initially).

By comparing the parameters of our system with the data obtained, we found (Drobyshevski *et al.*, 2001) that the ejection time of Auger electrons in ZnS(Ag) $t_{Aug} \sim 0.1$ ns, of nucleons from a nucleus $t_{ev} \lesssim 1$ ns, and the decay time of a daemon-containing proton $\Delta t_{ex} \approx 1$ μs. The fact that $A_n$ of the nuclei of our scintillator turned out to be optimum for our search played here a not insignificant role. The antisymmetric pattern of the $N_w(\Delta t)$ and $N_n(\Delta t)$ distributions for $|\Delta t| < 20$ μs suggests the existence of a SEECHO population with $20 < V < 35–40$ km/s (and/or of the Galactic disk population). The maximum at $20 < \Delta t < 40$ μs can be assigned to NEACHO daemons striking the Earth. A comparison of the April with May−June (2001) $N(\Delta t)$ distributions reveals maxima for $|\Delta t| > 50$ μs moving away from one another, which may originate from the GESCO population sinking under the Earth's surface in a time of 1−2 months (Drobyshevski *et al.*, 2001). The total daemon flux on the Earth $f_\oplus \sim 10^{-5}$ m$^{-2}$s$^{-1}$, a factor of ~300 less than our original (Drobyshevski, 1997) optimistic estimates.

3  SEMIANNUAL DAEMON-FLUX CYCLE AND ITS ORIGIN

Further exposure of the detector (to April, 2002) completed the year cycle and favoured our expectations that the Sun moves with respect to the daemon component of the Galactic disk and creates in the process a shadow and an anti-shadow of the SEECHO objects.

We may recall that we still do not discriminate the scintillations excited by a daemon crossing the detector from most of the other, background, scintillations. The only thing we do know is that the daemon must produce an HPS-type scintillation in the top scintillator. Our judgment of the daemon flux and of its characteristics draws on statistics only. And if there were no nuclear-active, strongly penetrating, low-velocity objects, the $N(\Delta t)$ distribution for $|\Delta t| > 1$ μs would be close to a constant level, because it would be produced by noncorrelated background events like the natural radiation, which is excited by cosmic rays as well. In particular, as pointed out in Sec. 2.2, the March, 2000, distribution $N(\Delta t)$ is, by the $c^2$ criterion, different from $N$ = const at a confidence level 1 - $a$ = 99.7%.

These data were processed practically manually. Presently, the processing is computerized. First, the monthly cosmic-ray signal statistics is used to determine the relative sensitivity of the scintillator-PM tube-amplifier systems for the top and bottom channels of each module. After this, the computer calculates a table whose cells correspond to a predetermined equal number of events to occur in each module $N_{mod}$ during a given time interval (month). The events chosen are such that their oscilloscopic trace amplitudes exceed their minimum values in the upper, $U_{1min}$, and bottom, $U_{2min}$, channels, with an additional condition imposed that a fixed ratio $w = U_{1min}/U_{2min}$ is the second parameter determining the table cell. Experience showed the meaningful ranges for our detector arrangements to be $50 \leq N_{mod} \leq 120$ and $1 \leq w \leq 7$. Each cell of the table contains the values of 1 - $a$ which are actually CL deviations of the $N(\Delta t)$, $N_w(\Delta t)$, $N_n(\Delta t)$, and $N_w(\Delta t) + N_n(-\Delta t)$ from a constant level. Already a cursory glance at the table reveals cells with maximum values of 1 - $a$. For $N(\Delta t)$, they usually lie in the ranges $N_{mod}$ = 80–90 and $w$ = 4–5.

Figure 2 presents a plot of 1 - $a$ for the detector with $h \approx 5$ mg/cm$^2$ at the upper 1 mm thick polystyrene screen (with $h \approx 3.5$ mm/cm$^2$ bottom 4 mm thick screen) obtained for the period from April, 2001, through April, 2002, which actually characterizes $N(\Delta t)$ (for $N_{mod}$ = 90 and $w$ = 4). It can be fitted by a sine function (with a correlation coefficient of 0.9, whose confidence level is 0.999) with a period of half a year and minima in the first decades of May and November (while the March, 2000, data correlate quite well with this relation, a certain misfit present stresses once more the importance of maintaining the scintillator density *h* at an optimum level). In our opinion, the results presented here are fairly unambiguous and would hardly allow any other interpretation except that the Earth crosses the solar shadow and anti-shadow in the SEECHO daemon flux, where parts of the NEACHOs crossing the Earth's orbit become also concentrated. This scenario does not contradict the opinion (e.g., Freese *et al*., 1988) that in the beginning of June the orbital velocity of the Earth adds to the Sun's velocity relative to the DM of the Galactic halo, and in the beginning of December, is subtracted from it. This means that the Earth must cross the solar shadow in the DM population sometime in February−March.

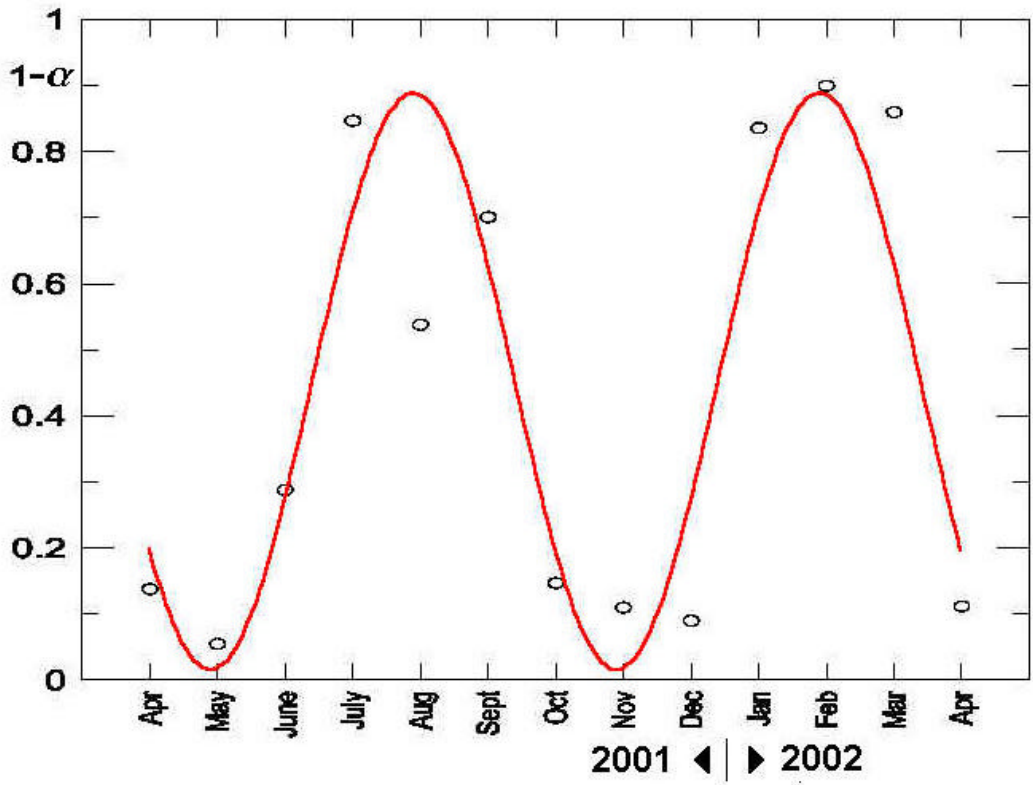

Figure 2. Monthly plot of 1 - $a$, the extent to which the $N(\Delta t)$ distribution deviates from the constant level produced by background events (see text for details).

## 4 CONCLUSION

Our original plans were aimed at detection of slow DM objects, which represent a population of the Solar system captured from the Galactic disk. Assuming them to be multiply negatively charged Planckian objects, daemons, we looked for a manifestation of strong nuclear effects, more specifically, simultaneous ejection of many scores of particles, internal conversion and/or Auger electrons and nucleons, in each interaction event. The crucial point in our ideology was the use of a supposedly fairly fast decay of the daemon-containing proton. Judging from the totality of the data obtained in our straightforward experiment, where it is simply unclear where an error could come from, we have succeeded in revealing the existence of the particles we have been looking for, which might perhaps be called superslow cosmic rays penetrating through the Sun and the Earth. It would appear close to inconceivable to assign these results to the action of the conventional cosmic rays. An analysis of the scintillation shape permitted us to understand a number of fine details in the daemon interaction with matter, i.e., with the main components of our detector. We have estimated the time of Auger electron emission in the daemon capture of a nucleus (∼0.1 ns), the time of evaporative de-excitation of the nucleus and of the time taken by the daemon to reach the ground state in its remainder (∼1 ns), and the time of the daemon-stimulated proton decay (∼1 μs). We revealed generically coupled daemon populations in various heliocentric orbits, as well as in geocentric orbits with perigees inside the Earth. The observed seasonal variations of the daemon flux correlate at a high confidence level (99%) with the semiannual cycle. They are due to the Earth's motion around the Sun, and of the Sun, relative to the DM of the Galactic disk, and can be accounted for in a noncontradictory way by the existence of daemon populations of several types.

Finally, the buildup of negative daemons in the Earth in an amount, as follows from our measurements, of ∼$10^{23}$, creates in it a kernel not more than a few cm in size (Drobyshevski, 2002a). Disintegration by the kernel daemons of Fe nuclei and of their protons is fully capable of explaining, besides many other phenomena that have been remaining unclear heretofore, the excess heat flux of ∼20 TW and the $^{3}$He flux emanating from the Earth's interior. Obviously enough, the daemons accumulated in the Sun (∼$2·10^{30}$) are capable of accounting, through proton disintegration, for a noticeable part of its luminosity and emission of the recently discovered neutrinos of nonelectron flavor. Note that an analysis of the processes occurring in the Sun's daemon kernel (with a radius of ∼10 cm) suggests the existence of negative daemons only (Drobyshevski, 2002c).

*References*

Birks, J.B. (1964) *The Theory and Practice of Scintillation Counting* Pergamon Press.
Drobyshevski, E.M. (1996) *Mon. Not. Roy. Astron. Soc.* **282**, 211.
Drobyshevski, E.M. (1997) in *Dark Matter in Astro- and Particle Physics* (H.V.Klapdor-Kleingrothaus and Y.Ramachers, eds.) World Scientific, p.417.
Drobyshevski, E.M. (2000a) *Mon. Not. Roy. Astron. Soc.* **311**, L1.
Drobyshevski, E.M. (2000b) *Phys. Atom. Nuclei* **63**, 1037.
Drobyshevski, E.M. (2000c) astro-ph/0008020.

Drobyshevski, E.M. (2002a) *Lunar & Planet. Sci. Conf.* (Houston, USA) **33**, Abstract 1119, 2 pp.; astro-ph/0111042.
Drobyshevski, E.M. (2002b) *Astron. Astrophys. Trans.* **21**(1) (in the press).
Drobyshevski, E.M. (2002c) astro-ph/0205353.
Drobyshevski, E.M., Beloborodyy, M.V., Kurakin, R.O., Latypov, V.G. and Pelepelin, K.A. (2001) *Astron. Astrophys. Trans.* (submitted to); Preprint PhTI-1753, St.Petersburg; astro-ph/0108231.
Freese, K., Friemann, J. and Gould, A. (1988) *Phys. Rev.* **D37,** 3388.
Markov, M.A. (1965) *Progr. Theor. Phys.* **Suppl., Extra Numb.,** 85.